\documentstyle[psfig,apjfonts,emulateapj]{article}

\slugcomment{Accepted Version (April 26, 1999)}

\lefthead{KASPI \& NETZER}

\righthead{MODELING VARIABLE EMISSION LINES IN ACTIVE GALACTIC NUCLEI:}

\def\gtorder{\mathrel{\raise.3ex\hbox{$>$}\mkern-14mu
                \lower0.6ex\hbox{$\sim$}}}
\def\ltorder{\mathrel{\raise.3ex\hbox{$<$}\mkern-14mu
                \lower0.6ex\hbox{$\sim$}}}

\def\CIV{C\,{\sc iv}\,}
\def\CIII{C\,{\sc iii}\,]\,}
\def\Lya{Ly$\alpha$}
\def\HeII{He\,{\sc ii}\,}
\def\MgII{Mg\,{\sc ii}\,}
\def\Si{Si\,{\sc iii}\,]\,}

\begin{document}

\title{Modeling Variable Emission Lines in AGNs: \\
Method and Application to NGC~5548}
\author{Shai Kaspi and Hagai Netzer}
\affil{School of Physics and Astronomy and the Wise Observatory, The
Raymond and Beverly Sackler Faculty of Exact Sciences, Tel-Aviv
University, Tel-Aviv 69978, Israel ; shai@wise.tau.ac.il,
netzer@wise.tau.ac.il }

\begin{abstract}

We present a new scheme for modeling the broad line region in active
galactic nuclei (AGNs). It involves photoionization calculations of a
large number of clouds, in several pre-determined geometries, and a
comparison of the calculated line intensities with observed emission
line light curves. Fitting several observed light curves simultaneously
provides strong constraints on model parameters such as the run of
density and column density across the nucleus, the shape of the
ionizing continuum, and the radial distribution of the emission line
clouds.

When applying the model to the Seyfert 1 galaxy NGC~5548, we were able
to reconstruct the light curves of four ultraviolet emission-lines, in
time and in absolute flux. This has not been achieved by any previous
work. We argue that the Balmer lines light curves, and possibly also
the \MgII$\lambda$2798 light curve, cannot be tested in this scheme
because of the limitations of present-day photoionization codes. Our
fit procedure can be used to rule out models where the particle density
scales as $r^{-2}$, where $r$ is the distance from the central source.
The best models are those where the density scales as $r^{-1}$ or
$r^{-1.5}$. We can place a lower limit on the column density at a
distance of 1 ld, of $N_{col}(r=1)\gtorder 10^{23}$ cm$^{-2}$ and limit
the particle density to be in the range of $10^{12.5}>N(r=1)>10^{11}$
cm$^{-3}$.

We have also tested the idea that the spectral energy distribution
(SED) of the ionizing continuum is changing with continuum luminosity.
None of the variable-shape SED tried resulted in real improvement over
a constant SED case although models with harder continuum during phases
of higher luminosity seem to fit better the observed spectrum.
Reddening and/or different composition seem to play a minor role, at
least to the extent tested in this work.

\end{abstract}

\keywords{
line: formation ---
galaxies: active --- 
galaxies: nuclei --- 
galaxies: Seyfert --- 
galaxies: emission lines --- 
galaxies: individual (NGC~5548)}

\section{Introduction}

Broad emission-Line Regions (BLRs) in Active Galactic Nuclei (AGNs)
have been the subject of extensive studies for more than two decades.
These regions are not spatially resolved and information about their
size, and the gas distribution, can only be obtained via reverberation
mapping. Understanding such regions requires a combination of detailed
observations with sophisticated theoretical models.

The extensive monitoring campaigns, of the last decade, focused on the
broad line emission, variability, and response of the broad emission
lines to the continuum variations in more than a dozen AGNs (see
Peterson 1993 and Netzer \& Peterson 1997, for reviews). A major result
is that in all well studied cases, the BLR is stratified with a typical
size of order of light weeks in Seyfert galaxies and light months in
quasars. Present day modeling of such stratified BLRs leaves much to be
desired. While time-averaged spectra of AGNs can reasonably be
explained by photoionization models (Netzer 1990 and references
therein), there is a lack of detailed models fitting to the actual gas
distribution, and motion, as deduced from the reverberation studies.

This paper investigates the physical conditions in the line emitting
gas of a well studied AGN by incorporating modern photoionization
calculations into a detailed geometrical model. Our approach, that
requires the simultaneous fit of several emission line light curves,
enables us to constraint several important properties of the BLR, such
as the gas density, column density and level of ionization. The model
is rather complex and is, therefore, limited to a spherical BLR with no
attempt to reconstruct the velocity field. In \S~2 we present the
model, in \S~3 we apply it to the Seyfert 1 galaxy NGC~5548, and in
\S~4 we discuss the new results in view of past studies.

\section{The Model}

\subsection{Motivation, choice of target and previous work}
\label{target}

Years of intensive AGN monitoring resulted in a wealth of information
on the BLR response to continuum variations. Excellent data sets are
now available for half a dozen sources and less complete sets on a
dozen or so more. Unfortunately, theoretical understanding lags behind
and there are few, if any, systematic attempts to produce complete
models for the objects in questions. Most theoretical efforts have been
limited to modeling of only one or two emission lines. They did not
consider real limitations such as the gas density and column density.
Our goal is to reconstruct {\em all the observed} light curves. This
will enable us to set stringent limits on the physical parameters of
the gas. More specifically, the aim is to deduce the run of density,
column density and cloud distribution across the BLR. We will
demonstrate that the time dependent relative and absolute line
intensities, and their relationship to the continuum fluctuations,
leave little freedom in such models. In particular, the density and
column density are restricted to a rather limited range.

We wish to use as many observational constraints as possible and,
therefore, limit ourselves to the best available data sets. We need
both recombination lines, to constraint the covering factor, and
collisionally excited lines to set limits on the run of density and
temperature. Data sets including both high and low ionization lines are
preferred since, in this way, we can set limits on the run of
ionization parameter as a function of distance from the continuum
source. Most importantly, we chose to rely as little as possible on
Balmer line intensities since, as explained below, their calculated
intensities are subjected to large theoretical uncertainties. Thus, the
chosen data set must include a large enough number of UV lines and a
well sampled, preferably large amplitude variable continuum. The only
information that will not be used in this work is line profiles.

Given the above limitations, there are only 4 data sets that we
consider suitable: 1. NGC~4151 - see Crenshaw et al. (1996) and
references therein 2. NGC~5548 (Clavel et al. 1991) 3. NGC~3783
(Reichert et al. 1994) 4. NGC 7469 (Wanders et al. 1997). Out of those
we chose to model the 1989 UV data of NGC 5548. This data set is
unique in terms of its relatively large line and continuum variations,
the {\em combination} of long duration with intensive regular 4-day
sampling, and complete information on all strong UV lines (see however
comments about \MgII$\lambda$2798 in the discussion section). Other
advantages of this light curve are discussed below.

In modeling the gas distribution in the nucleus, we distinguish between
direct and indirect methods. Direct methods are those where one guesses
a gas distribution and calculates the resulting line light curves by
putting, through this geometry, the observed (or deduced) continuum
light curve. The result is compared with the observed emission-line
light curves. Indirect methods attempt to obtain the gas distribution
by computing transfer functions. This procedure is somewhat ill-defined
since it produces responsivity maps, rather than mass-distribution
maps. It therefore requires the additional confirmation that the
so-obtained responsivity maps are consistent with photoionization
calculations. There have been several attempts to produce transfer
functions for various lines in NGC~5548, we are not aware of any
successful mapping that is consistent with full photoionization
calculations (see also Maoz 1994).

Several previous attempts to model the BLR in NGC~5548 differ from the
present study in several important ways. The first systematic attempt
to reconstruct the BLR is by Krolik et al. (1991). These authors used
photoionization calculations to find the parameters that best fit the
mean observed line rations. In parallel, they have used the Maximum
Entropy Method to reconstruct several transfer functions. Based on
these transfer functions, they modeled a spherically symmetric BLR with
locally isotropic line emitting clouds. The BLR is divided into two
distinct zones: one emitting the high-ionization lines (column density
of $\sim 10^{22}$ cm$^{-2}$ and ionization parameter of 0.3) and the
other emitting the low-ionization lines (column density of $\sim
10^{23}$ cm$^{-2}$ and ionization parameter of 0.1).

O'Brien, Goad, \& Gondhalekar (1994) have combined photoionization and
reverberation-mapping calculations in a way similar to the present
study (\S~\ref{formalism}). Their work, and also the one by P\'{e}rez,
Robinson \& de la Funte (1992), focused on the shape of the transfer
function and the line responsivity in a few generic models. They did
not attempt a detailed reconstruction of a specific data set, and the
comparison with the observations includes {\em only time lags}, and not
the detailed shape of the light curves.

Shields, Ferland, \& Peterson (1995) used photoionization calculations
to examine the \CIV$\lambda$1549/\Lya \ line ratio in NGC~5548 and
other objects. They concluded that an optically thin non-variable
component contributes to the BLR high ionization lines. Bottorff et
al. (1997) presented a detailed kinematic model which they combined
with photoionization calculations. They applied this model to {\em one
emission line} (\CIV$\lambda$1549) in the spectrum of NGC~5548.

Dumont, Collin-Suffrin, \& Nazarova, (1998) modeled the BLR in NGC~5548
as a 3-zone region where the location of the various components were
determined by the {\it mean} line-to-continuum lags. These authors
assumed high column density ($\geq 10^{23}$ cm$^{-2}$) and photon
density (the ionization parameter multiplied by the hydrogen density
gives 10$^{9}$ to 10$^{10}$ cm$^{-3}$). Much of their conclusion
regarding the density and column density is based on the relative
strength of the Balmer lines which we consider to be unreliable for
reasons we describe below. Like most others, they did not compare the
observed shape of the emission line light curves with the prediction of
the model.

Finally, in a recent work, Goad and Koratkar (1998) re-examined the
1989 NGC~5548 data set and deduced a strong radial dependence for the
gas density ($N\propto r^{-2}$, with density in the range of
10$^{11.3}-10^{10.0}$ cm$^{-3}$). Here again, the main assumption is of
a two-zone model and there is no attempt to use the observed light
curves, only the measured time lags.

Evidently, there is a definite lack of modeling complete data sets,
i.e. the information provided by half a dozen emission line light
curves. The published works focus on either detailed photoionization
calculations or detailed geometrical modeling (i.e. using transfer
functions) and no study have combined the two in a satisfactory way.
Thus, no previous work has demonstrated that the assumed gas
distribution, and the calculated physical conditions, are indeed
compatible. Global consistency checks are certainly required.

Our work relies heavily on the direct approach. We prefer to avoid, as
much as possible, ill-defined transfer functions and unreliable mass
distributions. Instead we make a large number of initial assumptions
(``guesses'') and check them, one by one, against the complete data set
of NGC~5548. This makes the numerical procedure more time-consuming but
is more robust because of the direct relationship between the assumed
geometry and the resulting light-curves. It also allow us to assess, in
a more meaningful way, the statistical significance of our results.
 
\subsection{Formalism}
\label{formalism}

We follow the approach described by Netzer (1990) for a BLR consisting
of numerous small clouds and a point-like ionizing source. The
contribution of each cloud to the various emission lines is determined
by the physical conditions within the cloud and its distance from the
center. The main assumption is that the cloud pressure is controlled by
an external medium (magnetic field, hot gas, etc.) whose pressure is a
simple function of the radius, $P \propto r^{-s}$. For completeness, we
summarize below the main ingredients of the model.

Consider a spherical system of clouds around a central radiation source
and assume pure radial dependence of all properties. The hydrogen
number density, $N(r)$, assumed to be constant within each cloud, is
controlled by equilibrium with the external pressure. Neglecting the
small electron temperature variation we take this to be
\begin{equation}
N(r)\propto r^{-s} \ \ .
\label{den}
\end{equation}
Next, we define the cloud column density, $N_{col}$, by considering
spherical clouds, of radius $R_{c}(r)$. The mass of the individual
clouds is conserved, as they move in or out, but it is not necessarily
the same for all clouds, thus, $R_{c}^{3}(r)N(r)=const$. The cloud
mean (over the sphere) column density is thus
\begin{equation}
N_{col}(r)\propto R_{c}(r)N(r)\propto r^{-2/3s} \ \ ,
\label{colden}
\end{equation}
and the geometrical cross-section is
\begin{equation}
A_{c}(r)\propto R_{c}^{2}(r)\propto r^{2/3s} \ \ .
\end{equation}
(Note that the mass conservation assumption is only essential if we are
to avoid an additional free parameter that specifies the run of
$N_{col}$ with distance). Finally, the number of clouds per unit
volume is
\begin{equation}
n_{c}(r)\propto r^{-p} \ \ .
\end{equation}
In the numerical integration we consider only one type of clouds (i.e.
same size at the same $r$ for all clouds). Simple generalization of
this scheme can involve a local population of clouds having some size
distribution and the same $p$ for all sub-classes. This requires extra
normalization (see below) which we decided to avoid at this stage.

The clouds are illuminated by a central source whose ionizing
luminosity, $L(t)$, varies in time. Designating $\epsilon_{l}(r,L)$ as
the flux emitted by the cloud in a certain emission-line, $l$, per unit
projected surface area ($erg \ s^{-1} \ cm^{-2}$), we get the following
relation for the emission of a single cloud:
\begin{equation}
j_{c,l}(r,L)=A_{c}(r)\epsilon_{l}(r,L) \ \ .
\end{equation}
(where we note that in the actual calculations, we use a thin slab
geometry, hence the geometrical cross section is somewhat different
from $\pi R_{c}^{2}$). Assuming the system of clouds extends from
$r_{in}$ to $r_{out}$ we obtain cumulative line fluxes:
\begin{equation}
E_{l}\propto \int_{r_{in}}^{r_{out}}n_{c}(r)j_{c,l}(r,L)r^{2}dr \ \ .
\label{El}
\end{equation}

Having determined the properties of the emission line clouds, and
having assumed a spectral energy distribution (SED) for the ionizing
source, we now calculate $\epsilon_{l}(r,L)$ using a photoionization
code. We then follow the above formalism to obtain $E_l$. Since all
parameters have been fixed, there is no need to independently specify
the ionization parameter, $U(r)$ which, in our model, is given by
\begin{equation} 
U(r)=\frac{Q}{4\pi r^{2}N(r)c} \propto r^{s-2}
\label{ionp}
\end{equation} 
where Q is the number of hydrogen ionizing photons per second, and $c$
is the speed of light.

AGN emission is variable, causing the local emission-line flux to be a
function of time. We take this into account by calculating
$\epsilon_{l}(r,L(t))$ for the entire range of continuum luminosity
applicable to the source under discussion. The calculated {\it in situ}
line fluxes are, therefore, the result of integrating Eq.~\ref{El},
using, at each distance $r$ the ionizing flux obtained with the
ionizing luminosity $L(t-r/c)$. Observed fluxes further require the
accounting for the light travel time between the cloud and the
observer.

A model is specified by the source luminosity and SED, the radial
parameter $s$, and the normalization of the various free parameters.
These include $r_{in}, \ r_{out}$ and the density and column density at
a fiducial distance which we take to be one light-day (1 ld), i.e.
outside the range of interest (in all the cases presented here
$r_{in}>1$ ld). The comparison with observations further requires the
normalization of the total line fluxes and hence the integrated number
of clouds. We prefer to use the total covering factor $C(r_{out})$
which is obtained by integrating
\begin{equation}
dC(r)=A_{c}(r)n_{c}(r)dr\propto r^{2/3s-p}dr \ \ 
\label{cr}
\end{equation}
between $r_{in}$ and $r_{out}$. The normalization of $C(r_{out})$ is
achieved by comparing the observed and calculated flux of the emission
lines. The numbers quoted below refer to the total covering factor as
determined by taking the mean covering factor derived for the
individual lines. We do not consider cloud obscuration and the
calculations are limited to values of $C(r_{out})$ smaller than about
0.3. Considering all this, the total number of free parameters in the
model is seven.

Computing a single model is not a trivial task. The reason is that, as
explained, $\epsilon_{l}(r,L(t))$ is a function of both distance and
time. We have accomplished this by calculating large grids of
photoionization models covering the required range of density, column
density, and incident flux. The calculations were performed using
ION97, the 1997 version of the code ION (see Netzer 1996, and
references therein). The code includes a large number of atomic
transitions and solves for the level of ionization, and the gas
temperature, at each location, taking into account all optical depth
effects. The line and continuum transfer is calculated by using the
local escape probability formalism. The cloud geometry assumed is of a
thin shell. This is somewhat different from the spherical-cloud
assumption made here, but the uncertainty introduced is well below the
uncertainty of the escape probability method.

A serious problem in ION, and in similar codes like {\it Cloudy} by G.
Ferland (Ferland et al. 1998), is the treatment of the optically thick
Balmer lines. The local escape probability approximation used in such
codes is inadequate for these lines (Netzer 1990). Unfortunately,
there is no known cure for this problem within that formalism and we
have decided, therefore, not to consider Blamer lines in this work. The
calculated intensity of other optically thick lines produced in the
large, extended, low ionization zone of the cloud (e.g.,
\MgII$\lambda$2798 and many FeII lines) is also considered uncertain. A
further limitation is the anisotropy of line emission which can be
important when emission line light-curves are concerned (see Ferland et
al. 1992 and references therein). We have decided not to include this
effect since the local escape probability method cannot accurately take
it into account and since it depends on the cloud shape, which is
unknown. We caution that while codes like ION and {\em Cloudy}
calculate this anisotropy as part of their standard output, the
fundamental limitation of the local escape probability method makes
such numbers highly uncertain.

We have adopted a goodness of fit criteria to evaluate the model
results. This is based on a simple $\chi^{2}$ score which is evaluated
by comparing the observed and theoretical light cures of {\it all}
chosen emission lines, giving each line an equal weight. Thus, an
acceptable model must reproduce the correct line variability as well as
the observed line ratios. To the best of our knowledge, no previous AGN
model have posed such stringent conditions on the goodness of fit.

\section{Modeling the BLR in NGC 5548}
\label{consider}

The Seyfert 1 galaxy NGC 5548 is a subject of an intensive study for
more then a decade now. This includes optical spectroscopic monitoring
for 8 years (Peterson et al. 1999, and references therein), several
satellite monitoring campaigns in different wavelengths --- each with a
duration of several months (e.g., Nandra et al. 1993, Korista et al.
1995, Tagliaferri et al. 1996, Marshal et al. 1997), and many
complementary studies of those data sets (e.g., Clavel et al. 1992,
Maoz et al. 1993, Wanders \& Peterson 1996). This, together with the
reasons mentioned in \S~\ref{target}, makes NGC~5548 the best object
for testing our model. In this study we have concentrated on the 1989
$International \ Ultraviolet \ Explorer \ (IUE)$ campaign (Clavel et
al. 1991) and used the $\lambda$1337 continuum and five emission lines;
Ly$\alpha\lambda$1216, \CIV$\lambda$1549, \CIII$\lambda$1909,
\HeII$\lambda$1640, and \MgII$\lambda$2798.

\begin{figure}
\centerline{\epsfxsize=3.4646in\epsfbox{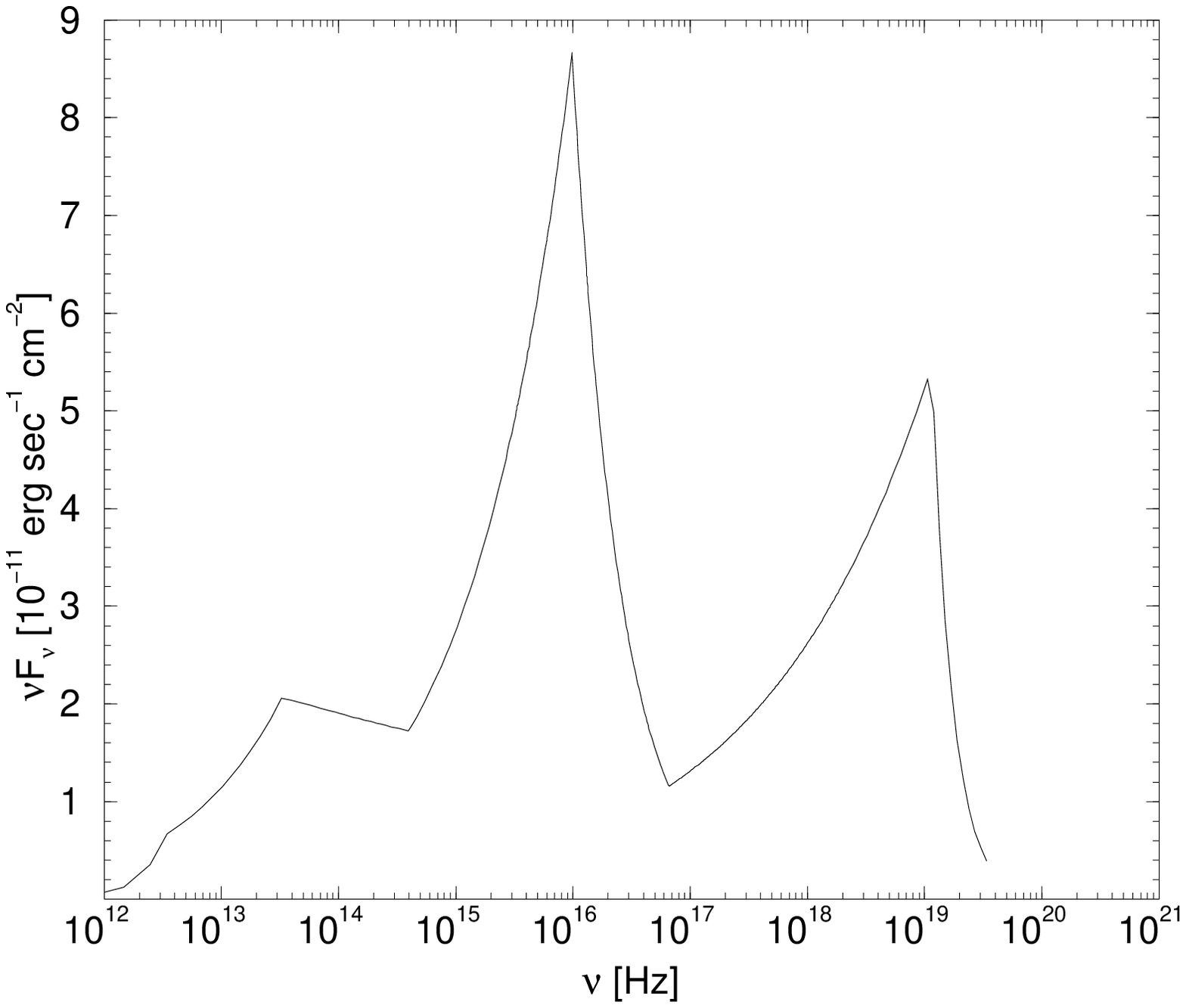}}
\caption{The spectral Energy Distribution of NGC~5548 used in this
study. The continuum is normalized to agree with Dumont et al. (1998).
\label{sed}}
\end{figure}

\begin{figure}
\centerline{\epsfxsize=3.4646in\epsfbox{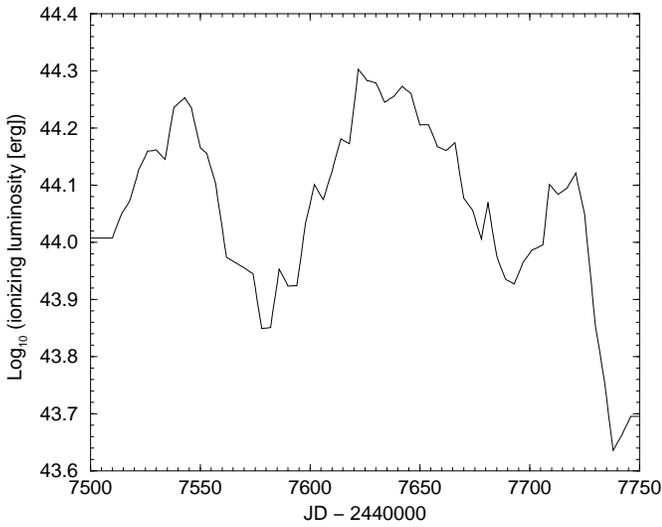}}
\caption{Interpolated ionizing continuum light curve based on the
Clavel et al. (1991) continuum and our assumed SED.
\label{con}}
\end{figure}

On top of the general considerations described in the previous section,
there are several constraints pertaining to NGC~5548:
\begin{enumerate}
\item 
The 1989 $IUE$ spectra, and the following HST observations (Korista et
al. 1995), clearly show that the \CIII$\lambda$1909 line is blended
with \Si$\lambda$1895. Clavel et al. (1991) have measured the combined
flux of both lines, without deblending (this is the flux listed in
their tables under the \CIII$\lambda$1909 header). This is also the
case in the Korista et al. (1995) paper. The observed ratio of
\CIII$\lambda$1909 to \Si$\lambda$1895 in the mean spectrum of
NGC~5548, is about 3:1. The line ratio, which is density dependent,
provides an important constraint on the density. We calculate the
intensities of the silicon and carbon lines and restrict ourselves to
those models producing a ratio consistent with the observations. The
diagrams and tables below gives the summed intensity under the title
\CIII$\lambda$1909.

\item 
Galactic and intrinsic reddening can affect the observed line ratios
and the assumed ionizing continuum. We address this in
\S~\ref{reddening}.

\item 
Another important consideration is the shape of the ionizing
continuum. The observed SED of NGC~5548 is reviewed by Dumont et al.
(1998) who discuss also various SEDs assumed in previous works. We have
chosen a range of possible SEDs, all in agreement with observations of
typical Seyfert 1 galaxies. Our optical-UV continuum is combined with
the observed ASCA continuum of the source assuming $\alpha_{ox}$=1.06
(see George et al. 1998 for a discussion of the X-ray spectrum of
NGC~5548). We have varied the assumed SED, as described in
\S~\ref{SEDch}, and tested the effect on the emission line spectrum.
All chosen SEDs are rather similar to the ones discussed in Dumont et
al. (1998). Our fiducial case is presented in Fig.~\ref{sed} on the
same scale as in Dumont et al. (1998).

\item 
Using the observed $\lambda$1337 continuum flux, a chosen cosmology
($H_0$=$75$ km s$^{-1}$Mpc$^{-1}$ and $q_0=0.5$), and the assumed SED,
we have estimated the time averaged ionizing luminosity of NGC~5548 to
be $\sim 10^{44}$ erg s$^{-1}$. The continuum flux at $\lambda$1337
varied by a factor of $\sim 4.5 $, during the $IUE$ campaign. Hence we
choose our grid to cover the ionizing luminosity range of $10^{43.6}$
to $10^{44.3}$ erg s$^{-1}$, with a step size corresponding to 0.1 in
$\log{L}$. The observed $\lambda$1337 light curve was transformed into
ionizing-luminosity light curve covering the entire range between these
values - see Fig.~\ref{con}. Considering the BLR geometry, we have
defined our grid of distances to cover the range of 1 to 100 lds, with
a step size of 0.02 in $\log{r}$.

\item 
We have adopted solar abundances for our standard model (see also
\S~\ref{gt} below).
\end{enumerate}

Model calculation proceed in two stages. First we produce
two-dimensional grids of $\epsilon_{l}(r,L(t))$ by calculating, for
each radial distance, all line emissivities for all ionizing
luminosities. Each grid represents a particular choice of the density
parameter $s$, and the density and column density at the fiducial
distance. We consider three density laws: $s=1$, 1.5, and 2, a density
at 1 ld of $\sim10^{11}$ to $\sim10^{13}$ (varying by a factor of
$\sim$2 from one grid to the next) and column densities at 1 ld of
$10^{22}$, $10^{23}$, and $10^{24}$. The choice of density, column
density, $s$ and the time-variable luminosity, completely specify the
ionization parameter at each location.

In the second stage we calculate theoretical light curves by
integrating Eq.~\ref{El}. For this we need to define $p$, $r_{in}$ and
$r_{out}$. We have considered models with $p=1$, 1.5 and 2. For each of
these we have varied $r_{in}$ and $r_{out}$ in order to minimize the
calculated $\chi^{2}$.

An important consideration is the error bar size for the first few
points. Obviously, we have no knowledge of the continuum behavior prior
to the beginning of the campaign. We have therefore decided to
extrapolated the continuum light curve, at a constant level, for all
JD~$<$~2447510. This lack of knowledge introduces a large uncertainty
in the first part of the line light-curves which is likely to be much
larger than the pure observational error. To partly cure this effect,
we have artificially increased all experimental error bars during the
first 100 days. We used a multiplication correction factor which is 3.0
for day 1 and decreases linearly with time, to 1.0, on day 100
(JD=2447600).

\subsection{A detailed example, $s=2$ models}
\label{example}

To illustrate our method, we present the results of a model with $s=2$
(i.e. $N(r)\propto r^{-2}$ and $N_{col}(r)\propto r^{-4/3}$) , $p=1.5$,
$N_{col}(r=1)=10^{24}$ cm$^{-2}$, and various densities. We note,
again, that what is referred to here as a {\it single model} covers, in
fact, a large range of density and column density. Thus, choosing
$N(r=1)=10^{12}$ cm$^{-3}$ and $N_{col}(r=1)=10^{24}$ cm$^{-2}$ we get
at 10 lds, $N=10^{10}$ cm$^{-3}$ and $N_{col}=10^{22.67}$ cm$^{-2}$,
and at 100 lds, $N=10^{8}$ cm$^{-3}$ and $N_{col}=10^{21.33}$
cm$^{-2}$.

\begin{figure}
\centerline{\epsfxsize=3.4646in\epsfbox{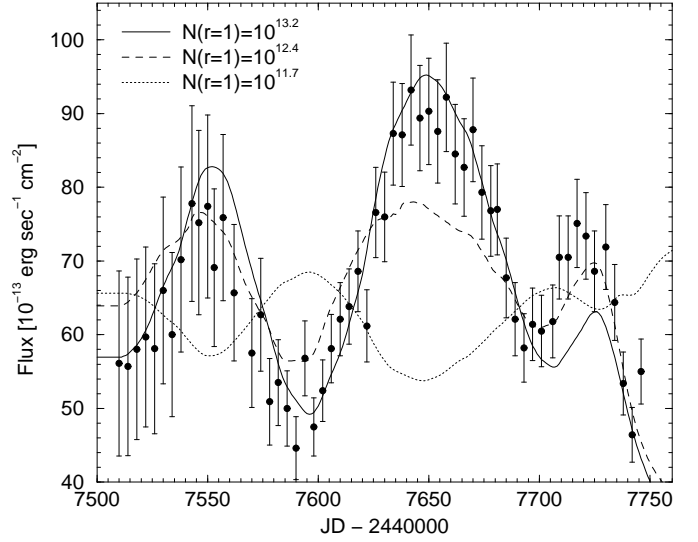}}
\caption{Calculated Ly$\alpha$ light curves for a model with $s=2$,
$r_{in}=3$ lds, $r_{out}=25$ lds $N_{col}(r=1)=10^{24}$, and various
densities --- $N(r=1)$, as marked. Dots with error bars represent the
observed light curve from Clavel et al. (1991).
\label{la_m2}}
\end{figure}

\begin{figure}
\centerline{\epsfxsize=3.4646in\epsfbox{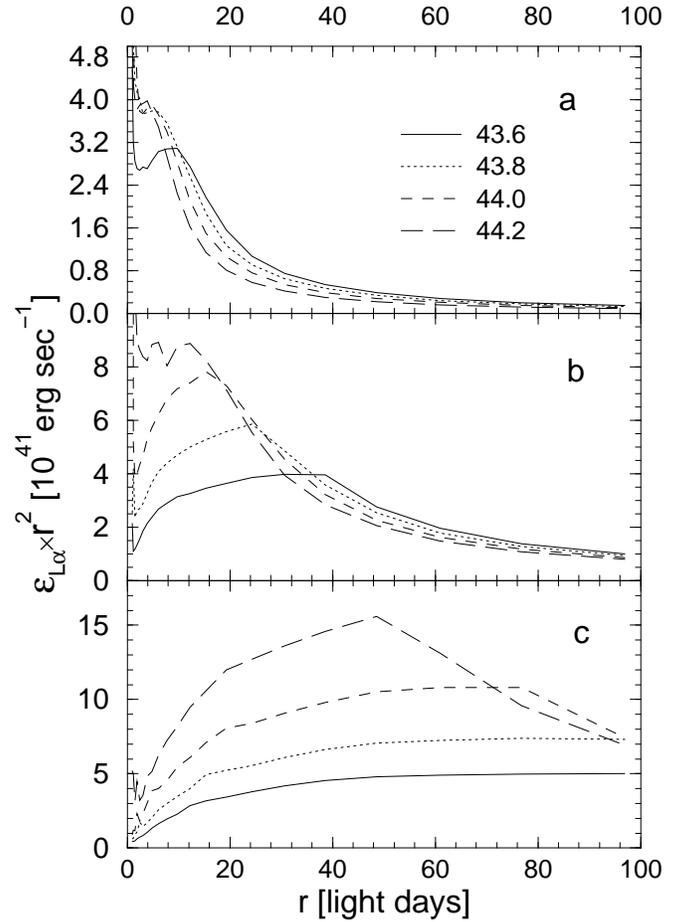}}
\caption{Ly$\alpha$ emission per unit solid angle versus $r$ for the
three densities shown in Fig.~\protect\ref{la_m2}. a:
$N(r=1)=10^{11.7}$ cm$^{-3}$. b: $N(r=1)=10^{12.4}$ cm$^{-3}$. c:
$N(r=1)=10^{13.2}$ cm$^{-3}$. For each model there are four different
ionizing luminosities, as marked.
\label{eps_m2}}
\end{figure}

Examining this case (Fig.~\ref{la_m2}) we note several obvious
features. First, for $N(r=1)\ltorder 10^{12.2}$ cm$^{-3}$, the emission
line response is reversed, i.e., increasing continuum luminosity
results in decreasing line luminosity. For example, while the ionizing
luminosity is increasing from JD$\sim$2447600 to JD$\sim$2447650
(Fig.~\ref{con}), the Ly$\alpha$ luminosity is decreasing (doted line
in Fig.~\ref{la_m2}), and the calculated light curve is a negative
image of the observed one. Further demonstration of this effect is
shown in Fig.~\ref{eps_m2}: Here we plot the line emission per unit
solid angle ($\epsilon_{Ly\alpha}(r,L)$$\times$$r^{2}$) versus $r$ for
four ionizing luminosities. For the lowest density $N(r=1)=10^{11.7}$
cm$^{-3}$ (Fig.~\ref{eps_m2}a)
$\epsilon_{Ly\alpha}(r,L)$$\times$$r^{2}$ is larger for lower ionizing
continuum for $r\gtorder 5$ lds, i.e, the emitted Ly$\alpha$ is getting
weaker when the ionizing luminosity is getting stronger. This effect is
seen also in all other emission lines. The reason for this behavior is
that for low density, the fast column density drop with distance causes
the clouds to become optically thin a few lds away from the center. For
lines other than Ly$\alpha$, the increase in ionizing flux causes the
gas to become too ionized to emit a significant amount in the line
under question. The decrease in Ly$\alpha$ is of different origin. This
line is mostly due to recombination and its intensity depends
exponentially on the Lyman continuum optical depth (which, in this
case, is of order unity) and linearly on the ionizing flux. Hence, the
decrease in the optical depth affects the line more than the increase
in ionizing flux. As evident from Figs.~\ref{la_m2} and \ref{eps_m2},
the reversed response disappears for $N(r=1)\gtorder 10^{12.2}$
cm$^{-3}$, resulting in a much better fit to the observations. This
shows that careful analysis of a single light-curve is already enough
to put limits on the BLR density.

\begin{figure}
\centerline{\epsfxsize=3.45in\epsfbox{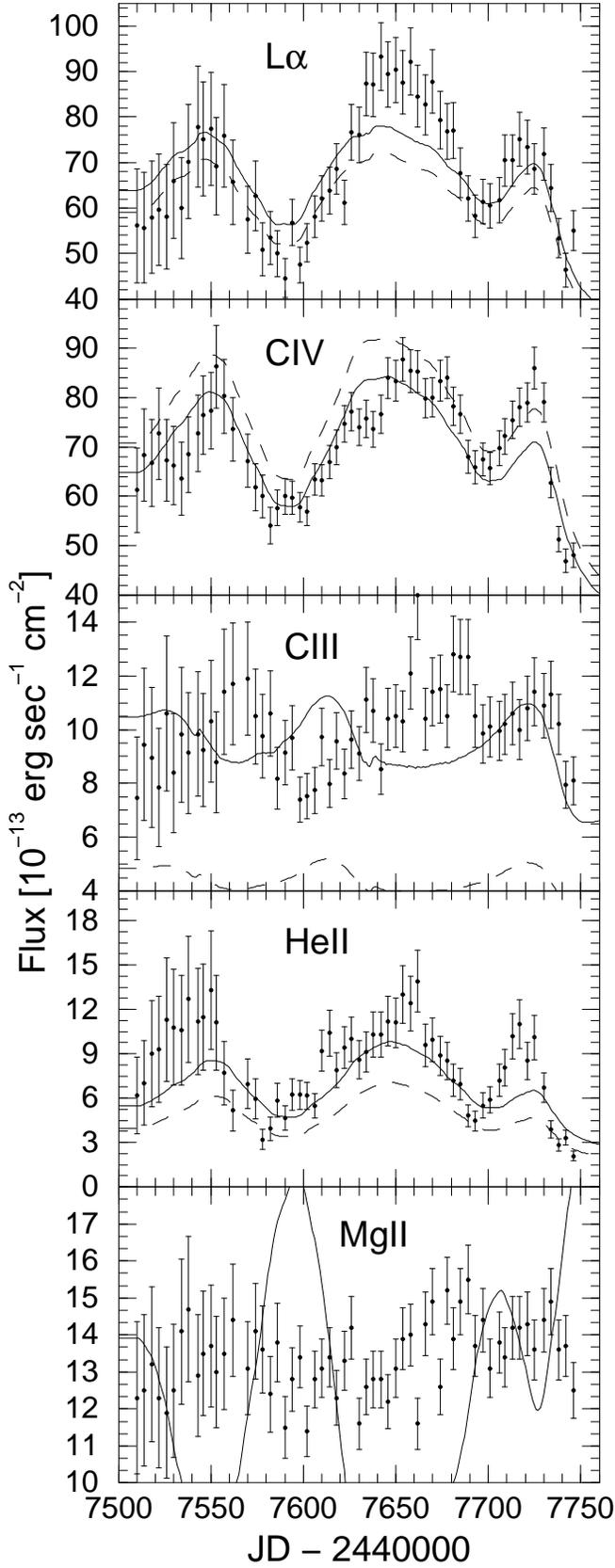}}
\caption{Simulated light-curves (continuous lines) compared with
observations for a model with $s=2$, $p=1.5$, $N(r=1)=10^{12.4}$
cm$^{-3}$, $N_{col}(r=1)=10^{24}$ cm$^{-2}$, $r_{in}=3$ lds, and
$r_{out}=25$ lds. The solid lines represent models with different
normalization factors for different lines. Dashed lines
represent models where the same normalization factor is used for all
lines (i.e. a single optimized covering factor)
\label{allr74}}
\end{figure}

\begin{figure}
\centerline{\epsfxsize=3.4646in\epsfbox{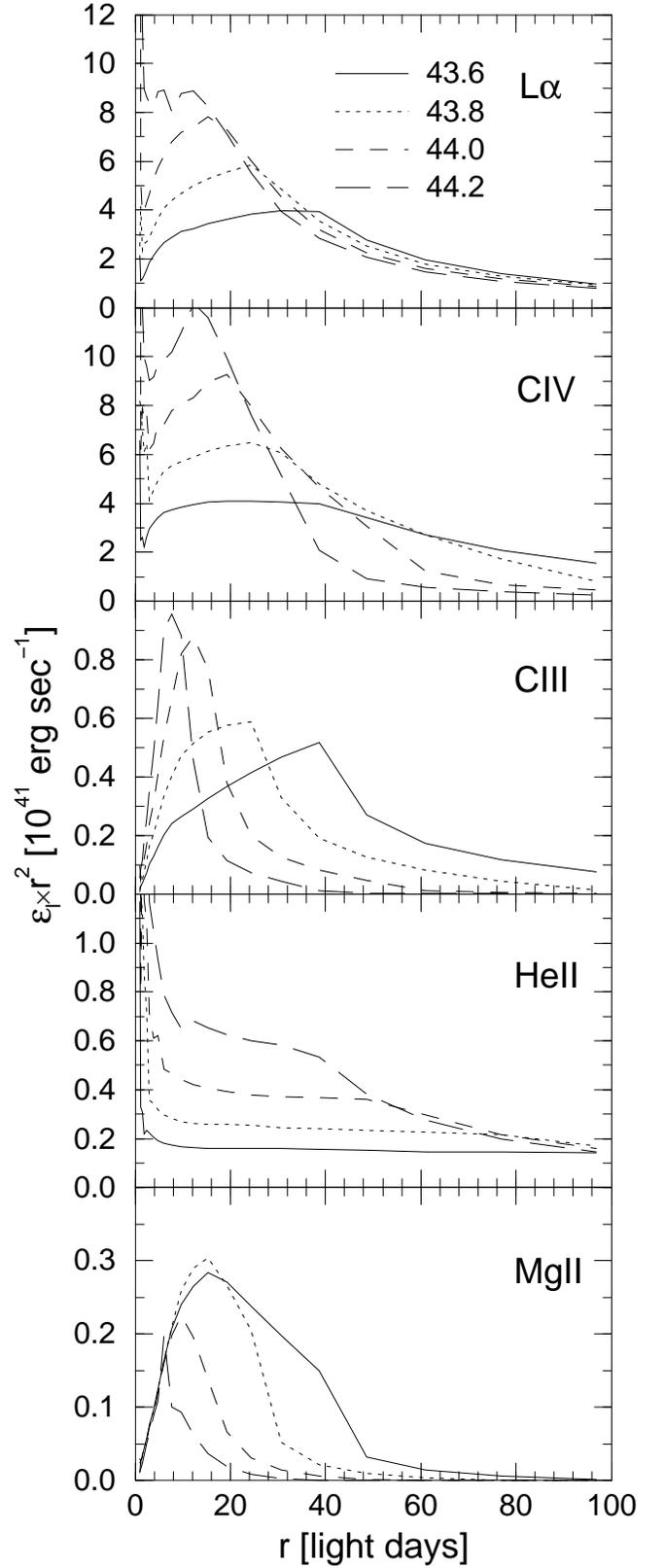}}
\caption{Line emission per unit solid angle versus $r$ for the 5
emission lines modeled in Fig.~\protect\ref{allr74}.
\label{r74}}
\end{figure}

While the simulated Ly$\alpha$ light curve, for the $s=2$ grid, nicely
fit the observed light curve with certain choice of density, $r_{in}$
and $r_{out}$, this geometry seem to do a poorer job for the other
emission lines. This is illustrated in Figs.~\ref{allr74} and \ref{r74}
that show that while calculated light curves of Ly$\alpha$, \CIV\ and
\HeII\ reasonably fit the observations, those for \CIII\ and
\MgII\ badly disagree with the data. Fig.~\ref{r74} shows the reason
for this discrepancy which is the positive response for Ly$\alpha$,
\HeII\ and \CIV\ up to about a distance of 30 lds, and the negative
response for \MgII\ and \CIII. Furthermore, the apparent good fit shown
by a solid line in Fig.~\ref{allr74} is misleading since it was
obtained by allowing different normalizations of Eq.~\ref{El} for the
different lines. When requiring the same covering factor for all
lines, there is no satisfactory solution (see the dashed line in that
diagram). The disagreement amounts to a factor 1.4 for \CIII\ and a
factor 12(!) for \MgII. For that model (dashed line in
Fig.~\ref{allr74}) the reduced $\chi^2$ score (for the 4 emission lines
excluding the \MgII line as explained in \S~\ref{gt}) is 9.8.

The above example demonstrates the difficulty in fitting,
simultaneously, line responses and line ratios. Most previous work
considered one line at a time and did not attempt full solution. Hence,
their seemingly satisfactory solution do not represent realistic
spectra. Our simulations show that no normalization or a choice of $p$
can cure this problem in the $s=2$ case. This is in contradiction to
the result of Goad \& Koratkar (1998) who favor an $s=2$ BLR. (see
further discussion of their model in \S~\ref{comp}). The underlying
reason for the failure of the $N\propto r^{-2}$ family of models is
that, in such cases the ionization parameter is constant throughout the
BLR. This is in contradiction to the observations that show different
lags for lines representing different levels of ionization.

\subsection{General trends}
\label{gt}

We have made a careful investigation of a large number of models. We
have considered cases with $s$ $=$ 1, 1.5, and 2, with $N(r=1)$ in the
range of $10^{11}$ to $10^{13}$ cm$^{-3}$, and three values of
$N_{col}(r=1)$, $10^{22}$, $10^{23}$ and $10^{24}$ cm$^{-2}$. Below is
a summary of the most important results.

We first note that as the column density increases from
$N_{col}(r=1)=10^{22}$ to $10^{24}$ cm$^{-2}$, so does the agreement
between the model and the data. Thus, models with
$N_{col}(r=1)=10^{22}$ cm$^{-2}$ give very poor fit (the line rations
are wrong and there is a reversed response in some lines) while models
with $N_{col}(r=1)\gtorder 10^{23}$ cm$^{-2}$ fit the data much
better. Current photoionization codes are limited to Compton thin gas,
i.e. they cannot reliably calculate cases with column densities larger
than about $10^{25}$. Thus we are unable to put upper limits on the
column density of the gas.

While large column density clouds are not sensitive to the exact
column, we have noted some changes that suggest that even the
$N_{col}(r=1)=10^{23}$ cm$^{-2}$ models are not thick enough. For
example, in some models with this column there is a noticeable
improvement in $\chi^{2}$ when increasing the BLR size (i.e. $r_{out}$)
which is not seen in the $N_{col}(r=1)=10^{24}$ cm$^{-2}$ models. It
indicates that the reversed response of optically thin cloud is present
even in $N_{col}(r=1)=10^{23}$ cm$^{-2}$ cases.

As shown in \S~\ref{example}, $s=2$ models are practically ruled out.
Regarding cases with $s=1.5$ and $s=1$, we have found that in both
cases there are lower and upper limits on the gas density. For
$N(r=1)\ltorder 10^{11}$ cm$^{-3}$, the $s=1.5$ models result in
reverse response (like the $s=2$ models) and the $s=1$ models produce
wrong line ratios (in particular in these low density models the
\CIII$\lambda$1909 to \Si$\lambda$1895 line ratio does not agree with
the observations, see \S~\ref{consider}). For $N(r=1)\gtorder
10^{12.5}$ cm$^{-3}$, the line ratios (e.g., \HeII/\Lya) in both $s$
cases do not agree with the observations. Hence, our study shows that
the BLR density in NGC~5548, is in the range of
$10^{12.5}>N(r=1)>10^{11}$ cm$^{-3}$. This is true for all models with
$N_{col}(r=1)\gtorder 10^{23}$ cm$^{-2}$.

A general trend found for all values of $s$ is that as $p$ increases
from 1 to 2 (i.e., more weight is given to clouds closer to the central
source) the amplitude of the modeled light curves is in better
agreement with the observations and so are most line ratios (an
exception is the Ly$\alpha$/\MgII\ line ratio - see discussion below).
Hence, models with higher $p$ are preferred.

A common problem in all models is the too weak Mg\,{\sc
ii}$\lambda$2798 line. While we do not have a complete explanation for
this, we suspect that it is due to one of two reasons: either the
transfer of this line is inaccurately treated, similar to the case of
the Balmer lines (discussed in \S~\ref{formalism}), or else it is
caused by the thousands of highly broadened Fe\,{\sc ii} lines, in that
part of the spectrum, that make the measured line intensity highly
uncertain (see for example the discussion in Wills, Netzer and Wills,
1985; Maoz et al. 1993). Below we discuss the possibility of enhanced
metallicity which also does not cure this problem. This is a definite
failure of our model and is the reason for excluding this line when
computing the $\chi^{2}$ and the normalization factor for
Eq.~\ref{El}.

Another common trend is the improved $\chi^2$ score for larger
$r_{out}$. In general, models with $r_{out}$ in the range 70 to 100 ld,
agree better with the observations compared with models with
$r_{out}<70$ lds. We also found that the models are not very sensitive
to the exact value of $r_{in}$ and the best value is 3 lds.

\begin{figure}
\centerline{\epsfxsize=3.2in\epsfbox{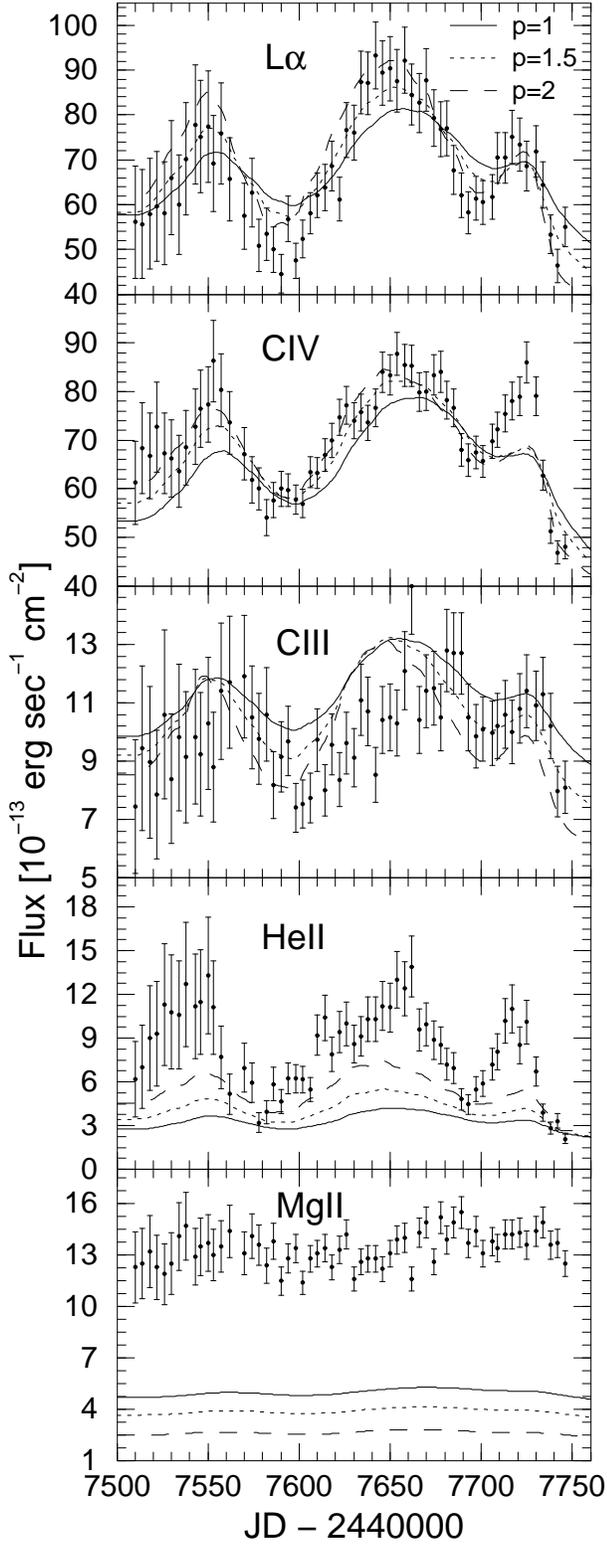}}
\caption{Simulated light-curves compared with observations for models
with $s=1$, $N(r=1)=10^{11.4}$ cm$^{-3}$, $N_{col}(r=1)=10^{24}$
cm$^{-2}$, $r_{in}=3$ lds and $r_{out}=100$ lds:  solid line - $p=1$ ;
dotted line - $p=1.5$ ; dashed line - $p=2$.  Note that $p=2$ model is,
by far, the best.
\label{r53n_4}}
\end{figure}

Considering the above trends and limitations, and using the $\chi^{2}$
score for the four chosen emission lines, we find that models with
$s=1$ best fit the observed spectra and models with $s=1.5$ give
somewhat inferior fits. An example of one of the best cases is a model
with $s=1$, $N_{col}(r=1)=10^{24}$ cm$^{-2}$, $N(r=1)=10^{11.4}$
cm$^{-3}$, $r_{in}=3$ lds and $r_{out}=100$ lds. The light curves
produced in this case are shown in Fig.~\ref{r53n_4} for various values
of $p$. The trend of improved fit with increasing $p$ is evident for
all lines except, as explained, for \MgII. The reduced $\chi^2$ scores
for the set of four emission lines are 4.5 for the $p=1$ model, 3.1 for
the $p=1.5$ model, and 2.2 for the $p=2$ model.

A note about the $\chi^2$ score is in order. Our best models are
characterized by $\chi^2 \sim 2$ which, statistically, is not very
significant. This is somewhat arbitrary since enlarging the estimated
errors by 50\% (which can be attributed to to systematic measurement
errors or model uncertainty) will result in $\chi^2 < 1$.

\subsection{Different composition}
\label{abundances}

As noted earlier, almost all previous studies used solar composition
material. In order to test the sensitivity of the results to the metal
abundance, we run our best fitted model (Fig.~\ref{r53n_4}) with all
metal abundances increased by a factor two. The results are generally
similar to the solar composition model, with somewhat larger $\chi^{2}$
score (slightly stronger \CIII\ and slightly weaker Ly$\alpha$ and
\CIV). In view of this we have not investigated other compositions.

\subsection{Reddening}
\label{reddening}

Reddening can affect line ratios as well as the apparent shape of the
continuum. Here we investigate the possibility that the intrinsic
continuum differs, in luminosity and SED, from the observed one and the
intrinsic line luminosities are larger than those observed. This would
have an effect on the gas distribution in the source since the amount
of continuum energy reprocessed by this gas, and the level of
ionization, are different under these conditions.

Galactic reddening toward NGC~5548 is discussed by Wamsteker et al.
(1990) who estimated it to be $E(B-V)=0.05$. These authors noted the
possibility of additional internal extinction which they have estimated
to be of the same order of the galactic reddening. Following this, we
have assumed an internal extinction identical to the galactic
extinction, i.e., total extinction of $E(B-V)=0.1$.

We have investigated the effect of reddening by modifying the SED in
our best fit model (section \S~\ref{gt} as shown in Fig.~\ref{r53n_4})
assuming the above $E(B-V)$. New line luminosities were obtained by
using extinction correction factors from Seaton (1979). Dereddening the
observed SED is more complicated since our adopted continuum is made of
a collection of power-laws. The adopted SED is presented in
Fig.~\ref{seds}e (dot-dashed line). It was normalized to fit the
reddened $\lambda$1337 light curve, in order to be consistent with the
observations (see explanation in \S~\ref{consider}).

The results of this model are essentially identical to those of the
unreddened case and are, therefore, not presented. This is not
surprising in view of the further tests that we have conducted that are
discussed below.

\subsection{Different SED}
\label{SEDch}

Several theoretical ideas (e.g. Binette et al. 1989, Clavel \&
Santos-Lleo 1990) as well as observational evidence (e.g., Romano \&
Peterson 1998 and references therein) point to a change in the SED
during continuum variation. This is expected since the shape of the
optical-UV continuum is known to undergo large changes (e.g. Wamsteker
et al. 1990, Clavel et al. 1991). Regarding the X-ray continuum
variations, these are of a different time scale, and different
amplitude, compared with the UV variations. It is therefore likely that
the shape of the UV-X-ray ionizing continuum is a complicated function
of the source luminosity. Unfortunately, there is little observational
study of such changes and we can only estimate the effect by testing
various possible SEDs and their effect on changes in the emission line
spectrum.

\begin{figure}
\centerline{\epsfxsize=3.4646in\epsfbox{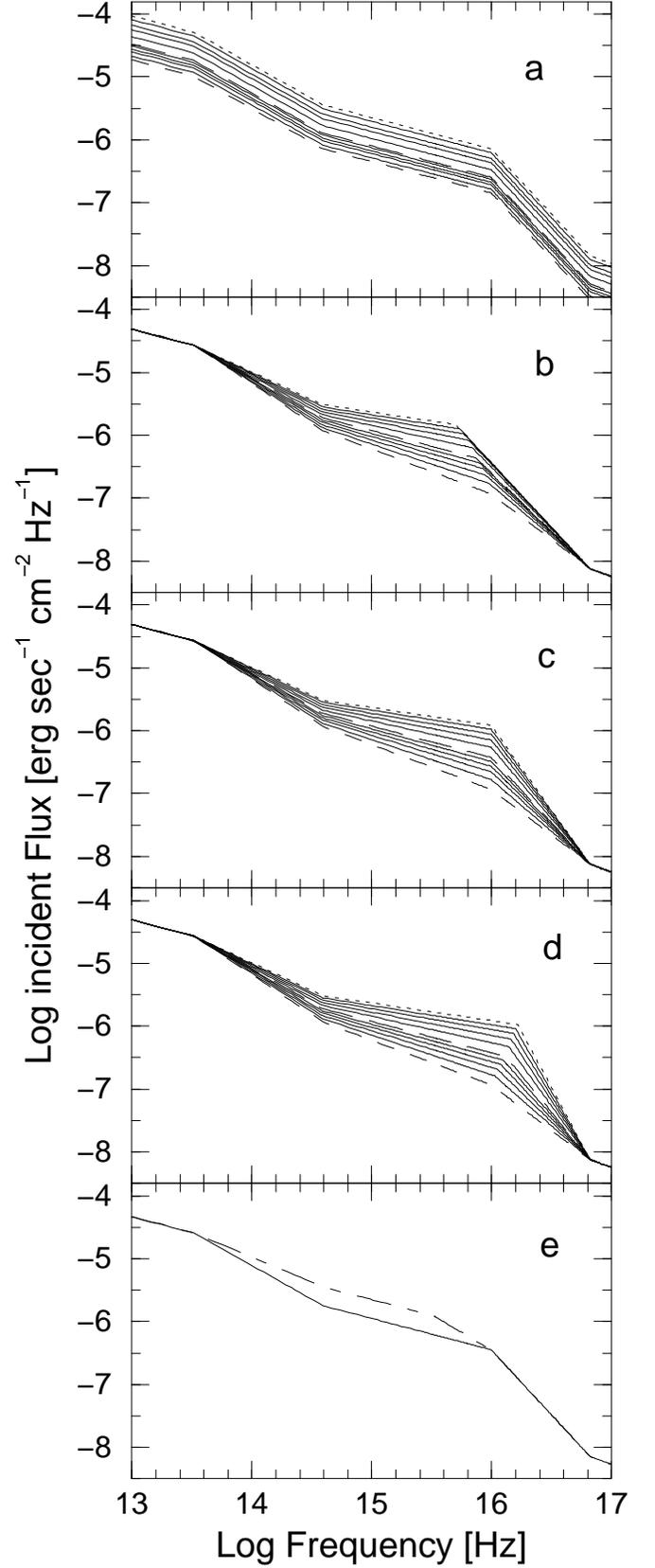}}
\caption{Different SEDs used in this work (see text for details).
We show the SED illuminating a cloud positioned at $r=10$ lds
with $N=10^{10.4}$ cm$^{-3}$ and $N_{col}=10^{23.3}$ cm$^{-2}$.
\label{seds}}
\end{figure}

We have carried out calculations with various possible schemes shown in
Fig.~\ref{seds}. The first case (Fig.~\ref{seds}a) is the standard SED
(the one used in all previous models, see Fig.~\ref{sed} and solid line
in Fig.~\ref{seds}e) shown here at 11 pre-chosen fluxes. Next we have
tried different shape SEDs. In doing so we are not interested in the
radio-FIR part of the continuum hence we did not alter the shape and
level of the spectrum below a frequency of 0.01 Ryd. We also keep the
flux above 20 Ryd constant since the {\it average} X-ray and
$\gamma$-ray flux does not change on time scales similar to the
optical--UV time scale (although it is changing very strongly on
shorter time scales). Keeping in mind the observed variation in the
optical-UV continuum slope, we assume that the flux change at 3 Ryd is
a factor 2 larger than the flux change at 0.12 Ryd (the 0.12 and 3 Ryd
are two of the energies used to define our SED). This is causing the
optical-UV continuum to become harder as the continuum luminosity
increases and is consistent with observations (Clavel et al. 1991). We
have normalized the SEDs such that the variable flux at 1337\AA
\ remains consistent with the observations. The different SEDs for this
scheme are presented in Fig.~\ref{seds}c. We have used them to compute
a new $\epsilon_{l}(r,L(t))$ grid for $s=1$ model and to produce light
curves for $p=1$,1.5 and 2. The agreement of these results with the
observations is not as good as our best fit, constant shape SED model
(\S~\ref{gt} and Fig.~\ref{r53n_4}) --- the $\chi^{2}$ score is about a
factor 2 worse.

The reason for the inferior fit is related to the relative number of
low and high energy ionizing photons. In particular, we attribute it to
the relative number of ionizing photons above 4 Ryd. In our original
model (Fig.~\ref{seds}a) the relative number of those photons is higher
than in the variable SED model (Fig.~\ref{seds}c). To verify this, we
have tested two other schemes. In the first scheme (Fig.~\ref{seds}b),
as the continuum flux increases the break point at 3 Ryd moves,
gradually, to {\em lower} frequencies (to 1.5 Ryd when the continuum
flux is highest). Thus, there are even fewer ionizing photons above 4
Ryd compared with the scheme shown in Fig.~\ref{seds}b. Indeed, the
resulting $\chi^{2}$ score is much worse (by about a factor 3.5)
compared with the constant SED scheme.

\begin{figure}
\centerline{\epsfxsize=3.4646in\epsfbox{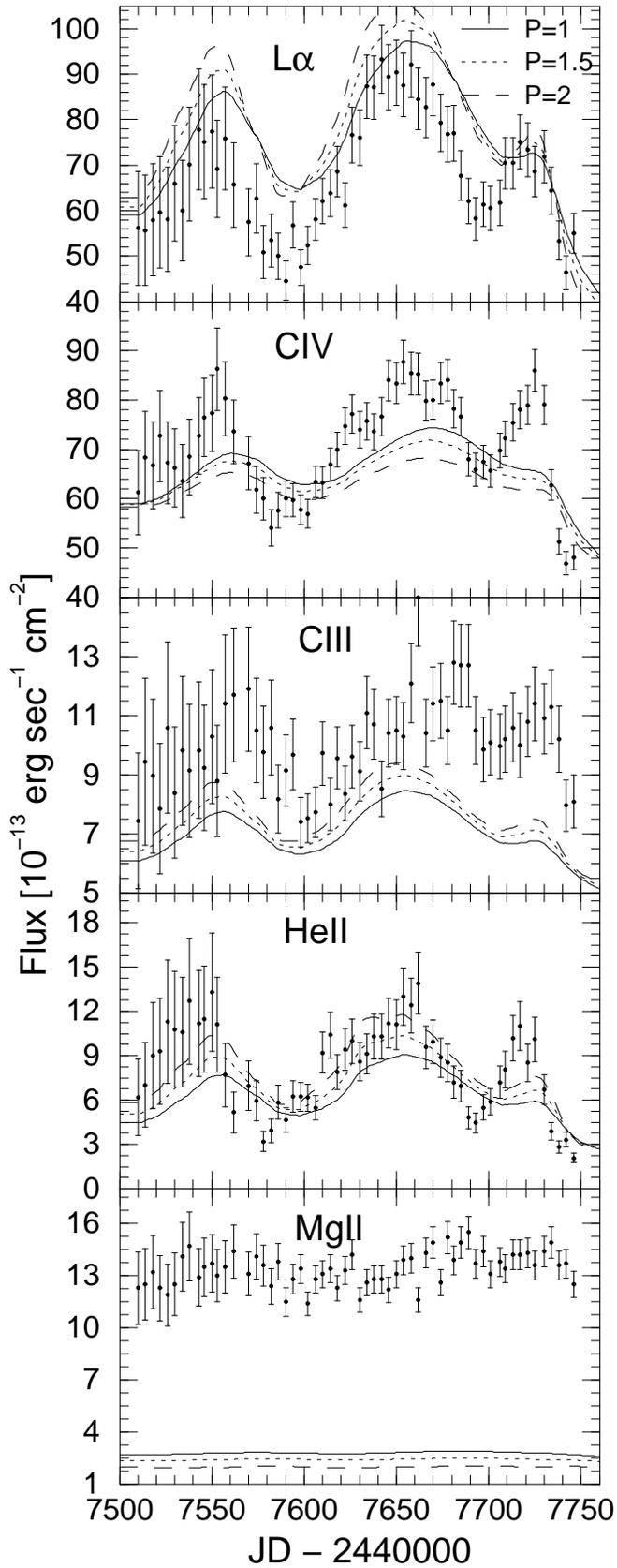}}
\caption{Same as Fig.~\protect\ref{r53n_4} for a model with variable
shape SED (Fig.~\protect\ref{seds}d).
\label{SEDsV}}
\end{figure}

In the second scheme (Fig.~\ref{seds}d), the break point moves to {\em
higher} frequencies as the continuum flux increases (from 3 Ryd when
the flux is lowest to 5 Ryd when the flux is highest). The relative
number of ionizing photons with energy larger than 4 Ryd increases with
continuum luminosity. The resulting $\chi^2$ is indeed better - see
Fig.~\ref{SEDsV}. In fact, the last scheme gives the very best fit of
all our models, to the \HeII\ line light curve. However, the fits to
the other lines (L$\alpha$, \CIV, and \CIII) are somewhat poorer and
the overall $\chi^{2}$ slightly worse (a factor 1.5) compared with the
constant SED case.

The above examples show that our model results are, indeed sensitive to
the assumed SED and its changes in time. Interestingly, the light curve
fits point to a change which is opposite to the one suggested by
Binette et al. (1989). We suspect that by scanning a large grid of
parameters with such variable SED we can improve our best result. This
is beyond of the scope of the present work.

\subsection{Covering factor}
\label{covering}

We have computed the covering factor by integrating Eq.~\ref{cr} as
described in \S~\ref{formalism}. For models with $s=2$ we find a
covering factors between 0.5 (for $p=2$) to 0.95 (for $p=1$). This is
much higher than predicted for typical AGNs but is in agreement with
the results of Goad \& Koratkar (1998) who find for their $N\propto
r^{-2}$ model a covering factor of $\sim$0.7. As explained, fitting the
light curves of several lines simultaneously, we can rule out models
with $s=2$. The too high covering factor seems to confirm our
conclusion.

Models with $s=1.5$ and $s=1$ result in smaller covering factors. For
our best model with $s=1$ we find a covering factor of 0.25 for $p=1$,
0.28 for $p=1.5$, and 0.30 for $p=2$. This is in good agreement with
previous estimates and is consistent with the no-obscuration assumption
of our model.

\section{Discussion and comparison with other studies}
\label{comp}

Several past models are detailed enough to allow meaningful 
comparison with our calculations.
\begin{enumerate}
\item
Our models are inconsistent with the $r^{-2}$ density law. This is in
disagreement with the study of Goad and Koratkar (1998) despite of the
good agreement with their range of acceptable densities and the lower
limit on the column density. Goad and Koratkar (1998) did not explore
a large parameter space, in particular, they only considered a limited
range of density laws. We suspect that their conclusion regarding $s=2$
models is related to the use of mean time lags, rather than detailed
fits to the observed light curves.

\item
Dumont et al. (1998) used a three zone model and reached several
conclusions based on time-averaged properties. They noted three
problems arising from their modeling: an energy budget problem, a line
ratio problem, and line variability problem. The first two are most
probably related to the use of Balmer line intensities as prime
elements in evaluating their model results. As explained, we have not
included the Balmer lines in our fit since we consider present day
photoionization models to be inadequate in this respect. Regarding the
line ratio, we find good agreement with observations for both high and
low ionization lines. We suggest that even a three-zone model, like the
one used by Dumont et al., is too simplified to account for the
continues gas distribution in NGC~5548.

\item
``Locally Optimally-emitting Clouds'' (LOCs) models (Baldwin et al.
1995, Korista et al. 1997) have been suggested to explain the broad
line spectrum of AGNs. The models assume clouds with a range of density
and column density at every point in the BLR. This is very different
from our assumption of a specific density and column density at each
radius. LOCs must be put into a real test by checking whether they
result in light curves that are in better agreement with the
observations, compared with the simpler models tested here.

There are two obvious ways to compare the two models. First, in LOC
models the incident central source flux ``selects'' clouds at certain
radii to be efficient emitters of certain lines while at other radii,
other lines are more efficiently produced. Thus, successful LOC models
of NGC~5548 should resemble ours, in their run of density and column
density for the most efficient emitting component. Second, future work
along the lines suggested here should test multi-component models
(i.e. various densities at each location) against the observed
light-curve. This will make them similar, in some respect, to LOC
models. Finally, we must comment that present day LOC models are too
general and do not contain full treatment of shielding and mixing of
the various coexisting components.

\item
Alexander and Netzer (1994; 1997) suggested that BLR clouds may be
bloated stars (BSs) with extended envelopes. In their work they fitted
the emission-line intensities, profiles and variability to the mean
observe AGN spectrum. One of the conclusion is that the density at the
edge of the BSs (the part emitting the lines) falls off like
r$^{-1.5}$, where r is the distance from the central source to the BS,
and the number density of the BSs falls off like r$^{-2}$. These two
trends are in good agreement with our preferred values of $s$ and $p$.
While the BS model is consistent with the mean time-lags measured in
intermediate luminosity AGNs, it remains to be seen whether it can fit,
in detail, the time dependent spectrum of objects like NGC~5548.
\end{enumerate}

\section{Summary}

We have presented a new scheme to model the BLR in AGNs. It combines
photoionization calculations with reverberation mapping and assumes
gradual changes of density and column density as a function of distance
from the central source. Using this scheme we were able to constraint
the physical conditions in the BLR and put limits on the important
physical parameters. The present work addresses only spherical BLRs.
Future work will hopefully generalize it to other geometries.

When applying the scheme to NGC~5548, we were able to reconstruct four
out of the five observed UV emission-line light curves. We found that a
large population of optically thin-clouds (due to low density and/or
column density) results in a reversed response to continuum variations.
Such models are therefore excluded. We also excluded models where
$N\propto r^{-2}$ and favor models where the density scales as
$r^{-1.5}$ or $r^{-1}$. We have placed a lower limit of
$N_{col}(r=1)\gtorder 10^{23}$ cm$^{-2}$ on the column density of the
clouds. We have used the results to place limits on the density of the
BLR. For NGC~5548, this must be in the range of
$10^{12.5}>N(r=1)>10^{11}$ cm$^{-3}$.

We have studied a range of possible SEDs, including one which resulted
from applying a reddening correction with $E(B-V)$=0.1. The one that
produces the best agreement with the observations is an SED where the
relative number of $E>4$ Ryd photons increases with the AGN
luminosity.

There are obvious limitations to this method and several ways to
improve it. For example, line beaming (anisotropy in the emission line
radiation pattern) must be considered. Unfortunately, similar to the
Balmer line problem, the radiation pattern cannot be accurately
calculated in present-day escape-probability-based codes.
Multi-component models, with a range of density and column density at
each radius (not necessarily similar to LOC models), must be tested
too. Finally, testing different BLR geometries and a denser parameter
space is highly desirable.

\acknowledgments 

We are grateful for constructive suggestions by the referee, B.M.
Peterson. We acknowledge financial support by the Israel Science
Foundation, and the Jake Adler Chair of Extragalactic Astronomy. S.K.
acknowledge financial support by the Colton Scholarships.



\twocolumn

\end{document}